\documentclass[journal,twoside,web]{ieeecolor}
\usepackage{jsen}
\usepackage{cite}
\usepackage{mathtools}
\usepackage{amsmath,amssymb,amsfonts}
\usepackage[verbose]{placeins}
\usepackage{algorithmic}
\usepackage{graphicx}
\usepackage{textcomp}
\usepackage{float}
\usepackage{tablefootnote}
\usepackage{graphicx}
\usepackage{wrapfig}
\def\BibTeX{{\rm B\kern-.05em{\sc i\kern-.025em b}\kern-.08em
    T\kern-.1667em\lower.7ex\hbox{E}\kern-.125emX}}
\definecolor{abstractbg}{rgb}{0.89804,0.94510,0.83137}
\begin{document}
\title{Piezoelectric PVDF-TrFE/ PET Energy Harvesters for Structural Health Monitoring (SHM) Applications}
\author{Berkay Kullukçu, Mohammad J. Bathaei,  Muhammad Awais, Hadi Mirzajani and Levent Beker
\thanks{Manuscript received XX, 2022; revised XX, 2022; accepted March XX, 2022. Date of publication XX, 2022; date of current version XX, 2022. The associate editor coordinating the review of this paper and approving it for publication was XXX. (Corresponding author: Levent Beker.)}
\thanks{B. Kullukçu is with the Mechanical Engineering Department, Koç University, Istanbul, Turkey (e-mail: bkullukcu15@ku.edu.tr). }
\thanks{M.Awais is with the Biomedical Engineering Department, Koç University, Istanbul, Turkey (e-mail: mawais21@ku.edu.tr).}
\thanks{J.Bathaei is with the Biomedical Engineering Department, Koç University, Istanbul, Turkey, (e-mail: sbathaei20@ku.edu.tr).}
\thanks{H.Mirzajani is with the Mechanical Engineering Department, Koç University, Istanbul, Turkey, (e-mail: hmirzajani@ku.edu.tr).}
\thanks{L.Beker is with the Mechanical Engineering Department, Koç University, Istanbul, Turkey, (e-mail: lbeker@ku.edu.tr).}
\thanks{Digital Object Identifier: XXXX}}

\IEEEtitleabstractindextext{%
\fcolorbox{abstractbg}{abstractbg}{%
\begin{minipage}{\textwidth}%
\begin{wrapfigure}[15]{r}{3.1in}%
\includegraphics[width=3in]{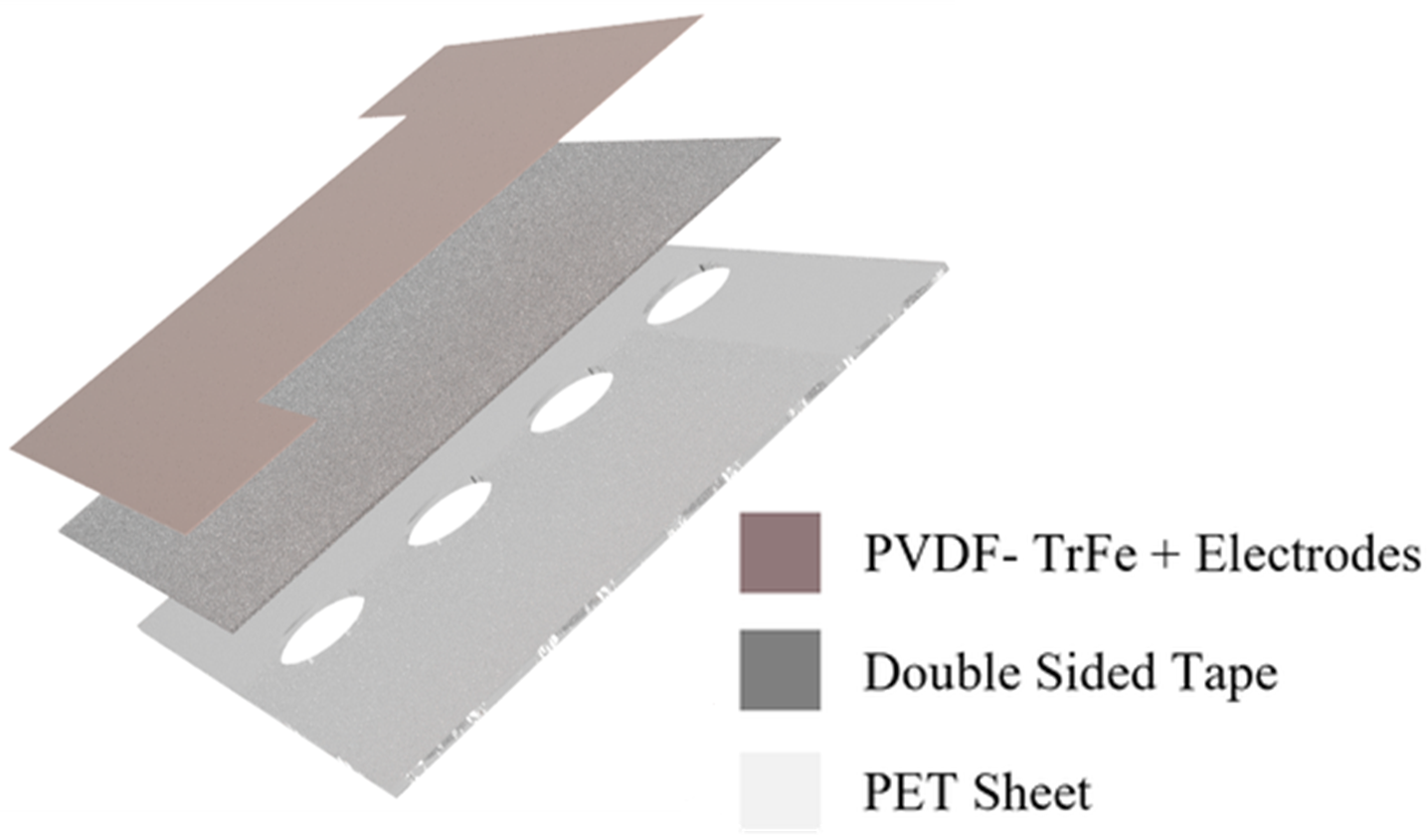}%
\end{wrapfigure}%
\begin{abstract}
This research describes a piezoelectric Poly(vinylidene fluoride-co-trifluoroethylene)/ Polyethylene Terephthalate energy harvester for structural health monitoring of wind turbines. The piezoelectric energy harvester was made of a polyvinylidene fluoride-trifluoroethylene (PVDF-TrFE) layer. In addition, PET sheets, double-sided micron-thick tapes, and PVDF sheets were used for device fabrication. In order to optimize the performance of the proposed device, the electrical and mechanical characteristics of PVDF-TrFE material were investigated. Two different piezoelectric energy harvester devices were designed, fabricated, and compared. A wind test setup was developed, and the proposed device was characterized by output voltage and current. Furthermore, a 104 nF capacitance was coupled to the proposed device to store the generated voltage and run a commercial pressure sensor. The generated output power of the harvester was sufficient to drive a MEMS pressure sensor with a maximum current and voltage of 291.34 µA and 937.01 mV, respectively, under wind pressure giving a power value of 272.99 µW.
\end{abstract}

\end{minipage}}}

\maketitle

\section{Introduction}
\label{sec:introduction}
\IEEEPARstart{T}{he} field of microelectronics has made tremendous advancements, resulting in the emergence of various electronic devices that have become commonplace in daily life.
The continual advancement of energy harvesting technologies has reduced these devices’ cost, size, weight, and power needs therefore, the formation of distributed environments, has been made feasible as a consequence of this. The existing energy storage capacity continues to be the major factor in determining the size, mass, and cost of modern electronic equipment. Batteries typically lead people to be worried about the inﬂuence that their disposal has on the environment as well as the feasibility of replacing them in systems.
Because of this problem, there seems to be a rise in interest in the development of energy harvesting technologies, that are predicted to resolve the issues that are brought about by the usage of energy sources in electronic devices [1].
Sensors, actuators, transmitters, receivers, and processors, all highly integrated intelligent electronic devices, have paved the way for various new life-improving applications. The development of technology known as microelectromechanical systems, or MEMS, has led to improvements in the functionality of electronics along with reductions in both size, cost, and amount of power they use [2]. Distributed systems were made possible by the ongoing trend toward smaller, less expensive equipment and advancements in wireless communication [3]. These networks have the potential to be utilized for various purposes. Examples of potential applications include detecting hazardous chemical compounds in high-traffic areas and measuring tire acceleration and pressure [4, 5]. Various industry professionals believe that very low wattage embedded electrical equipment will soon be an essential component of our day-to-day lives, performing various activities ranging from the automation of factories to our entertainment needs [6]. Despite this, the eﬀectiveness of these emerging technologies is dependent on the accessibility of reliable sources of energy. Cost, size, and duration are issues for energy sources with fixed capacities, such as batteries. This research is based on energy harvesting, an alternative approach to addressing these challenges. 
The most recent iteration of structural health monitoring (SHM) systems incorporate low power sensors and wireless communication components fueled by energy harvesters’ outputs [7]. Energy can be extracted from the immediate environment to power autonomous sensor systems [8]. SHM integration into wind energy systems is a lucrative application for demonstrating SHM technology in general [9]. The components that must be inspected are large and difficult to inspect. During operation, wind turbines are subjected to intense and complex wind loadings; therefore, evaluating their dynamic responses to wind loads is desirable due to the possibility of fatigue and the desire to better comprehend the turbine’s response [10]. In addition to fundamental condition assessment, monitoring may contribute to more efficient turbine operation. Wind energy harvesting has a limitation known as the” Betz Limit” (in practice, a multiplier of 0.593 (16/27) is applied to the maximum deliverable power). SHM can be performed with vibration-activated transducers because vibration is a common occurrence in nature. In recent years, there has been a great deal of interest in vibration energy harvesters employing piezoelectric transduction mechanisms due to their higher power density, ease of application, and scalability [11]. In situations with no fixed vibration, wind vibration powers piezoelectric harvesters. This thesis describes a piezoelectric energy harvester device dedicated to SHM in wind turbines by providing the necessary power for electronic devices through vibrational energy harvesting.
\section{Modeling of the Harvester Device}
In the electrical domain, there is a capacitor; in the mechanical
domain, there are spring and mass elements. Ideal transformer
elements are used to indicate domain coupling. The mechanical
domain damping caused by anchor loss and other factors is
insignificant. Hence it is not considered in the model. Table I
contains the model parameters.
The capacitance, C, is described in the electrical domain as a
conventional parallel plate capacitor having capacitance
defined by [12]:
\begin{table}
\caption{Values for Equivalent Circuit of PEH}
\label{table}
\setlength{\tabcolsep}{3pt}
\begin{tabular}{|p{35pt}|p{105pt}|p{75pt}|}
\hline
Symbol& 
Description& 
Units\\
\hline
C& 
Electrical Capacitance& 
F\\
$r_{pm}$& 
Radius of the top
electrode& 
m\\
$t_{pvdf}$& 
PVDF thickness& 
m\\
$\varphi(x_{\theta})$& 
Normal plate
deflection shape
function&
-\\
$I_m$& 
Piezoelectric coupling
integral& 
-\\
$I_e$& 
Strain energy integral& 
-\\
M& 
Piezoelectric bending
moment& 
${CVs^2}/{kg}$\\
$e_{31,f}$& 
Transverse
piezoelectric
coefficient& 
C/N\\
$V_{in}$& 
Applied voltage& 
V\\
z& 
Neutral plane& 
m\\
$t_{k},z_k,E_k$& 
Layer thickness, the
center axis of each
layer, Youngs modulus& 
m,m,Pa\\
$v_k$& 
Poisson's ratio& 
-\\
D,$h_{k}$& 
Flexural rigidity, the
distance to the top of
each layer& 
$Pam^4$,m\\
$\mu_{eff}$& 
Effective mass per unit&
-\\
$rho_{k}$&
Area, density& 
$m^2,kg/m^3$\\
$k_{m}$&
Stiffness& 
$kg/s^2$\\
$\eta$&
Electromechanical
coupling& 
Cm/N\\
$f_{n}$&
Resonance frequency&
Hz\\
$\lambda_{01}$&
Eigenvalue for the first
mode vibration&
-\\
$m_{m}$&
Total mass&
kg\\
$m_{d}$&
Modal mass&
kg\\
\hline
\end{tabular}
\label{tab1}
\end{table}

\begin{equation}C = {\varepsilon\pi{r_{pm}}^2}/{t_{pvdf}}.\label{eq}\end{equation}
where $\varepsilon$ the proportion of relative permittivity of
PVDF to the permittivity of a vacuum, $t_{pvdf}$ represents the
PVDF thickness, and $r_{pm}$ represents the radius of the top
electrode. With a normal plate deflection shape function of $\varphi$(x) {[}13{]}:

\begin{equation}
\label{eq4}
\varphi(x) = {(1-{x_\theta}^2)}^2.
\end{equation}
where $x_\theta$ is the radial coordinate in its normalized form. The axisymmetric plate deflection is  w($x_\theta$) = $w_0\varphi$(x), where $w_0$ is the static plate deflection at the clamped plate's center. One can calculate $I_m$, the piezoelectric coupling integral, and $I_e$, strain
energy integral by which are specified as {[}12{]}:
\begin{equation}
\begin{multlined}
\label{eq5}
I_m = 1/{(1-v)} \int_{0}^{\gamma} {((x_{\theta}d^{2}\varphi(x_\theta)}/{d{x_\theta}^2)}+\\{(d\varphi(x_\theta)}/{dx_\theta))dx_\theta}.
\end{multlined}
\end{equation}

\begin{equation}
\begin{multlined}
\label{eq6}
I_e =  \int_{0}^{1}((x_{\theta}{{d^{2}\varphi(x_\theta)}/{d{x_\theta}^2})^2}+\\ 2v{({d\varphi(x_\theta)}/{x_{\theta}dx_{\theta})}{(d^{2}\varphi(x_\theta)}/{d{x_\theta}^2})}+{(d\varphi(x_\theta)}/{dx_\theta))}d{x_{\theta}}.
\end{multlined}
\end{equation}

where v is the composite plate's effective Poisson's ratio, and M is the piezoelectric bending moment provided by [14]:
\begin{equation}
\label{eq7}
M = -e_{31,f}V_{in}z.
\end{equation}
The transverse piezoelectric coefficient is $e_{31,f}$. The applied
voltage is $V_{in}$. The distance between the active PVDF layer's
mid-plane and the neutral plane is z. It is worth noting that the
piezoelectric coupling integral $I_m$ is computed solely over the
electrode region, but the strain energy integral $I_e$ is computed
across the whole radius. The location of the composite plate
structure's neutral plane z is defined by:
\begin{equation}
\label{eq8}
z = {\sum_{k=1}^{3} {(t_{k}z_{k}E_{k}}/{1-v_{k}^2)}}/{\sum_{k=1}^{3}(t_{k}E_{k}}/{1-v_{k}^2)}.
\end{equation}
Counting from the bottom, the subscripts denote the thin plate
layers. The Young's modulus is $E_{k}$ the Poisson's ratio is $v_{k}$, and the center axis of each layer is $z_{k}$. Similarly, the composite
plate's flexural rigidity, D [15], and effective mass per unit area, $\mu_{eff}$:
\begin{equation}
\label{eq9}
z = \sum_{k=1}^{3} 1/3( {{{(h_{k}-z)}^3}-{(h_{k-1}-z)}^3)}/{({1-{v_{k}}^2}/E_k)}.
\end{equation}
\begin{equation}
\label{eq10}
\mu_{eff}=(1/D)\sum_{k=1}^{3} \rho_{k}t_k.
\end{equation}
$\rho_{k}$ is the density and $h_k$ represents the height of each layer.
is the distance to the top of each layer.
One gets when solving Eq. 11 for the mechanical compliance $1/k_m$:
\begin{equation}
\label{eq11}
1/k_m = {r^2}/{2{\pi}DI_e}.
\end{equation}
The following equation for the electromechanical coupling ratio
is found by solving:
\begin{equation}
\label{eq12}
\eta = 2{\pi}I_{m}e_{31,f}z.
\end{equation}
The circular plate formula is utilized [13] to get the natural
frequency:
\begin{equation}
\label{eq13}
f_{n}=({{{\lambda_{01}}/r)}^2}\sqrt{D/{\mu_{eff}}}.
\end{equation}
where $\lambda_{01}$ is the corresponds to the (01) vibration mode eigenvalue. As the plates will be vibrating in their first eigenfrequencies, $\lambda_{01}$ was used throughout the equations. 
The total mass of the composite disk, $m_d$ and the shape function
is used to determine the modal mass:
\begin{equation}
\label{eq14}
m_{d}= ({\rho_{\rho\rho}t_{\rho\rho}+\rho_{be}t_{be}+\rho_{a\rho}t_{a\rho})\pi\\r^2\\+{\rho_{te}t_{te}\pi\\{r_{te}}^2}}.
\end{equation}
\begin{equation}
\label{eq15}
m_{m}= m_{d}*2\int_{0}^{1} {{\varphi(x_\theta)}^2}x_{\theta}dx_{\theta}.
\end{equation}
where the densities of the distinct layers are $\rho_{\rho\rho}, \rho_{be},\rho_{a\rho}$ and $\rho_{te}$ respectively. The finite element model (FEM) was developed to measure the
voltage output of the device. An axisymmetric geometry was run under FEM simulation. Fig. 2 shows \textsc{COMSOL}$^{\circledR}$
FEM simulation results of the circular plate.
\section{Experiments}
PVDF-TrFe was selected as the actuation material for the
piezoelectric harvester. Table 2 shows the superiority of PVDF-TrFE compared with other widely used piezoelectric materials
in terms of voltage output KPI.
\begin{table}
\caption{Piezoelectric Material Comparison}
\label{table}
\setlength{\tabcolsep}{3pt}
\begin{tabular}{|p{35pt}|p{35pt}|p{35pt}|p{35pt}|p{35pt}|p{35pt}|}
\hline
Parameter&
$BaTiO_3$&
Quartz&
PZT4&
PZT5H&
PVDF-TrFe \\
\hline
$d_{33}$ [$10^{-12}$]& 
180 [18]& 
2.3 [22]&
289 [17]&
593 [16]&
-38 [17]\\
$\rho [{kg}/{m^3}]$& 
6010 [20]& 
2620 [23]&
7500 [24]&
7500 [19]&
1800 [21]\\
${\varepsilon^x}/{\varepsilon_0}$&
1557 [18]&
5 [22]&
1300 [22]&
3400 [21]&
9.4 [21]\\
$g_{33}$ {[}\(10^{-3}\){]} &
  13.06 &
  51.98 &
  25.12 &
  19.71 &
  456.79 \\
\hline
\end{tabular}
\label{tab1}
\end{table}
PVDF-TrFe material characterizations were determined in terms of thickness, pyroelectric, and ferroelectric measurements. In order to characterize the material, it is necessary to first use 2wt$\%$ PVDFTrFE in acetone as solvent (Piezotech FC25). The specimen is then placed on a magnetic stirrer for 1 hour at 80°C. The solution was then filtered to make it more homogenous. The filtered solution was spin-coated on glass substrates sweeping between 750 RPM to 5000 RPM. After this operation, the solution was annealed for 1 hour at 140 °C. Fig. 1 shows the thickness of the deposited layer for different RPMs.
\begin{figure}[H]
\centerline{\includegraphics[width=\columnwidth]{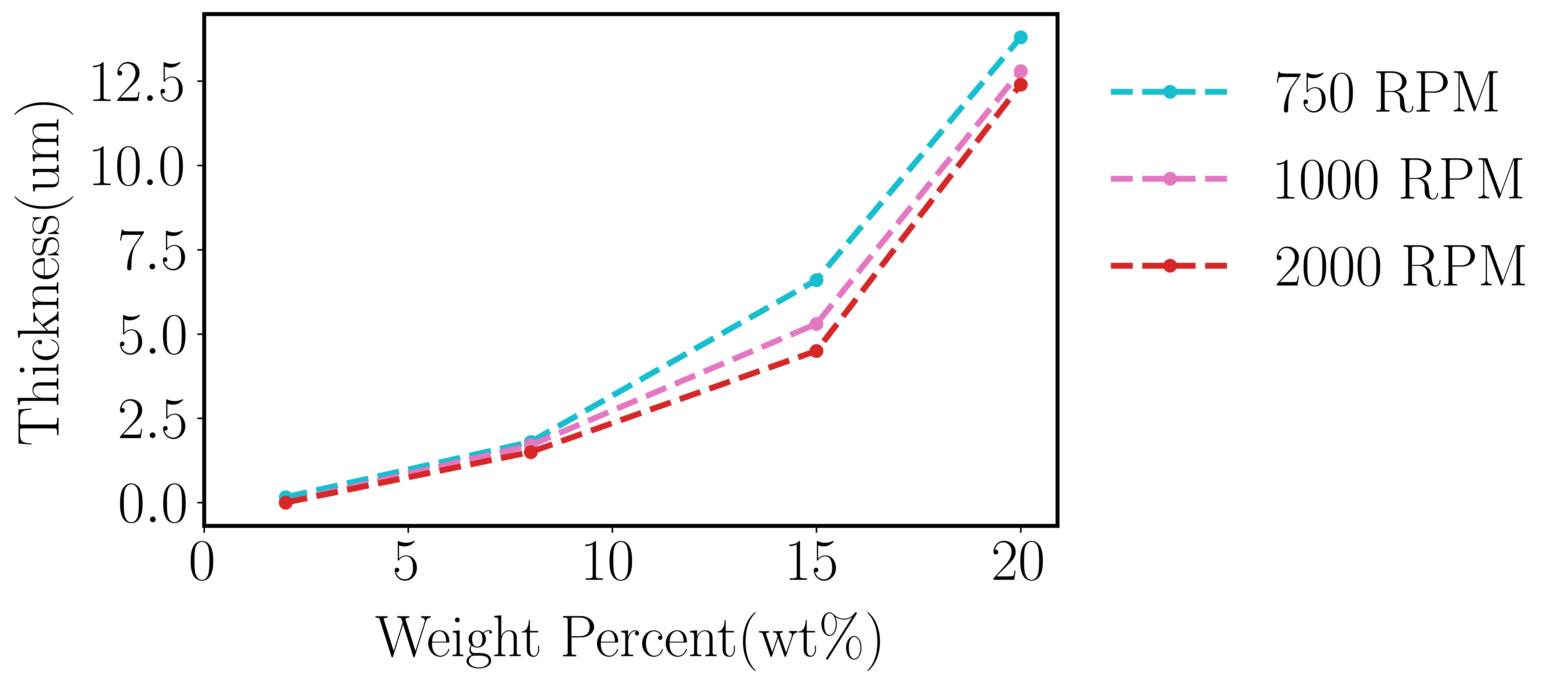}}
\caption{Profilometer thickness measurements were taken with
Dektak XT for different spin-coated specimens.}
\label{fig1}
\end{figure}

Film formation was not desirable in 5000 RPM spin-coated film as short-circuit tests for this specimen failed for most regions when it was spin-coated on ITO glass—indicating high nonuniformity or porosity of the surface. Film formation was better in 2000 RPM than in 5000 RPM spin-coated film. However, the short circuit test failed on the edges of ITO glass. A 33wt$\%$ PVDF-TrFe (FC25 Arkema Piezotech) solution was too viscous to be squeezed using the pipettes. In conclusion, lower RPMs ($<$ 2000) resulted in better film uniformity, but acetone seems not to be a suitable solvent for PVDF-TrFE due to its highly volatile nature. Poling was done using the high voltage DC power supply. An additional 100 V was applied for 1 µm thickness. A Sawyer-Tower Circuit was built for conducting ferroelectric measurements with the specimens. Fig. 2 shows the result of these ferroelectric measurements conducted with the circuit.
\begin{figure}[H]
\centerline{\includegraphics[width=\columnwidth]{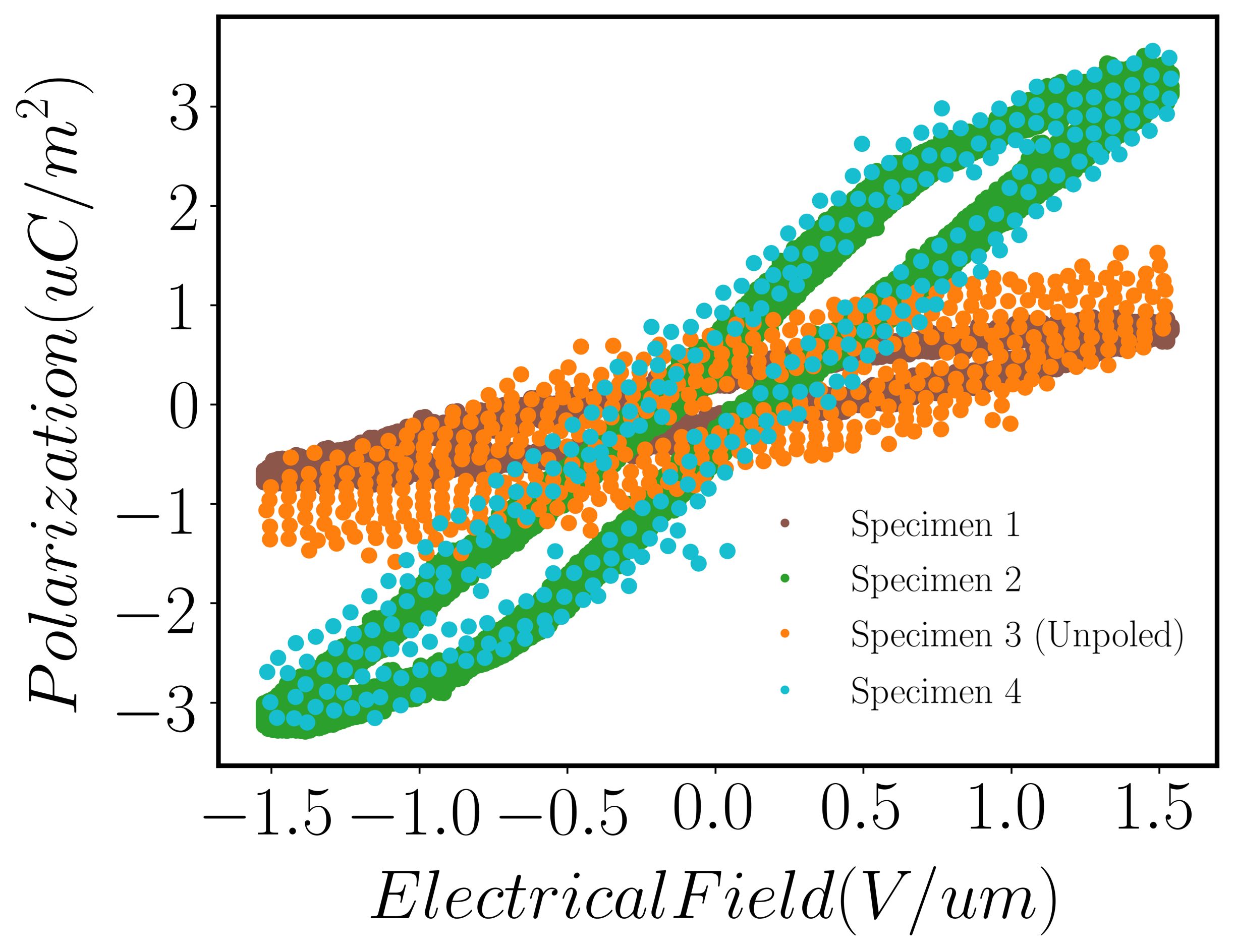}}
\caption{Polarization results for 4 different test specimens are
given, and measurements are done using triangular waves with
20 $V_{pp}$.}
\label{fig1}
\end{figure}

PVDF-TrFE has a dual-phase microstructure consisting of alpha and beta phases where the percentage of the beta phase determines the stiffness and crystallinity of the film. Fig. 3 represents the test setup used for stress-strain curve plotting and represents the stress-strain curve in the elastic region for the semi-crystalline PVDF-TrFE copolymer.
\begin{figure}[H]
\centerline{\includegraphics[width=0.9\columnwidth]{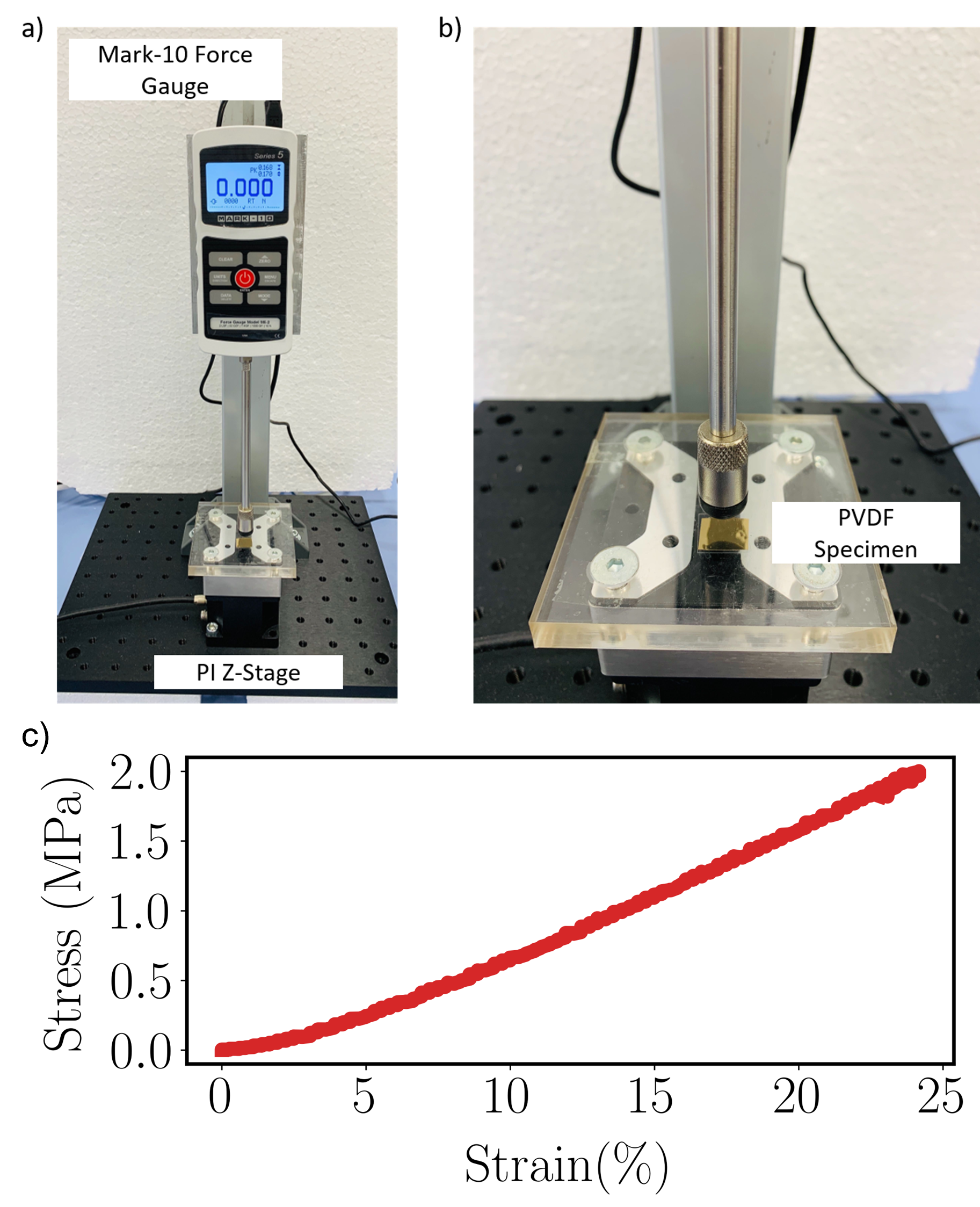}}
\caption{a) Compressive tests were done using a force gauge and PI L-306 Compact Precision Z-Stage b) Metal coated PVDF specimen was used for the experiment c) Stress-strain plot of the material resembles a linear curve in strain levels as high as 25$\%$.}
\label{fig1}
\end{figure}

According to Fig. 3c, the polymer has a Young Modulus of 0.87 GPa, proving its robustness. In addition to the high stiffness of the polymer, PVDF-TrFE is a suitable candidate for energy harvesting applications where flexibility and mechanical robustness are required [25]. According to equation (1) where $g_{33}$ is piezoelectric voltage coefficient, d is thickness, and s is applied stress, the voltage output of a piezoelectric film is directly proportional to the applied stress on the film. Fig. 4 demonstrates the flexibility of the film under a microscope and Fig. 5 shows an optical image of a harvester prototype.

\begin{figure}[H]
\centerline{\includegraphics[width=0.9\columnwidth]{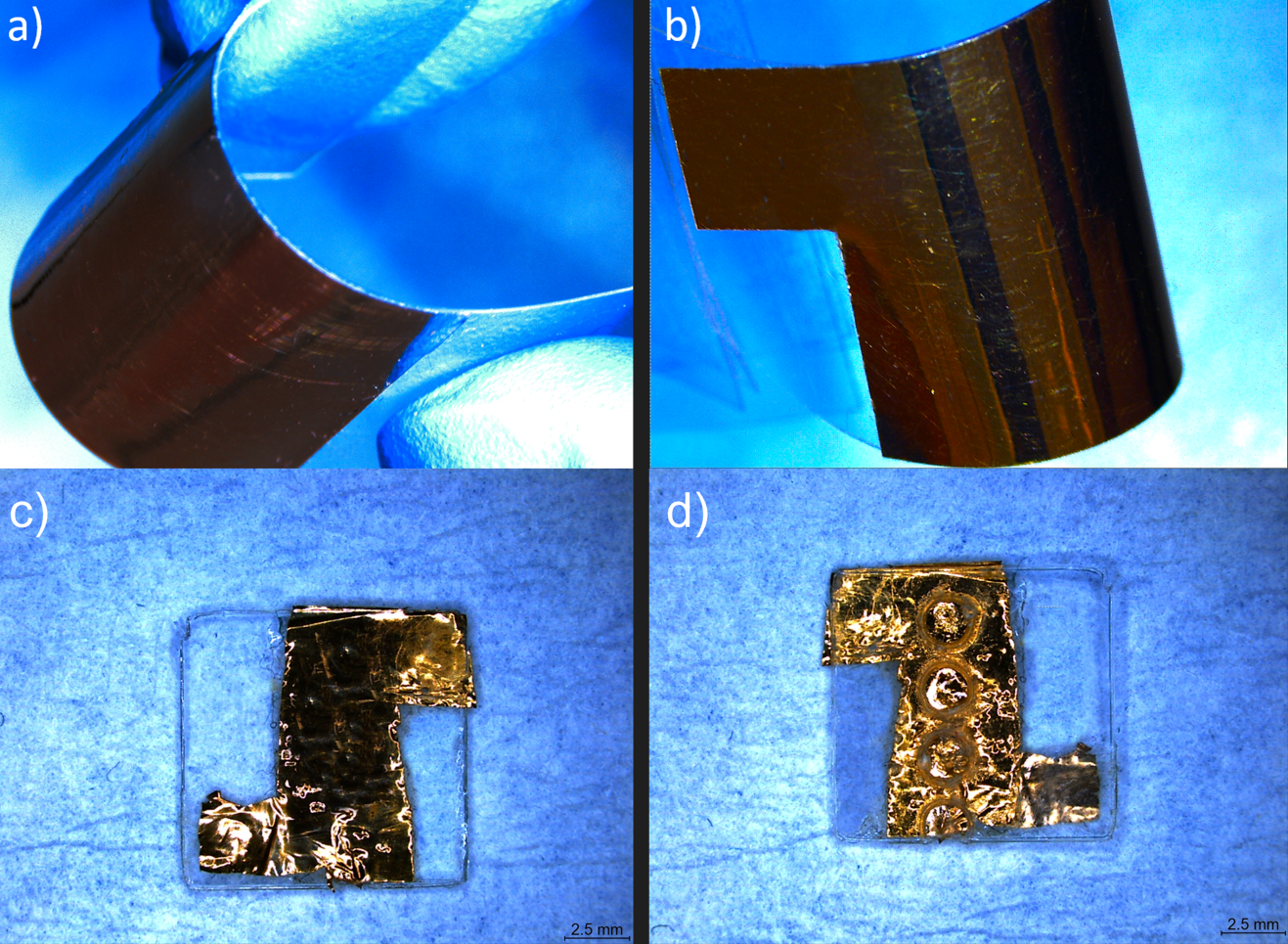}}
\caption{a) Optical image representing the flexibility of the harvester b) PET layer was used as a substrate for the piezoelectric layer to bend without going into a tear, permanent deformation c) Optical image of the fabricated sensor from the top view (First Design) d) Bottom view of the same design.}
\label{fig1}
\end{figure}

\begin{figure}[H]
\centerline{\includegraphics[width=\columnwidth]{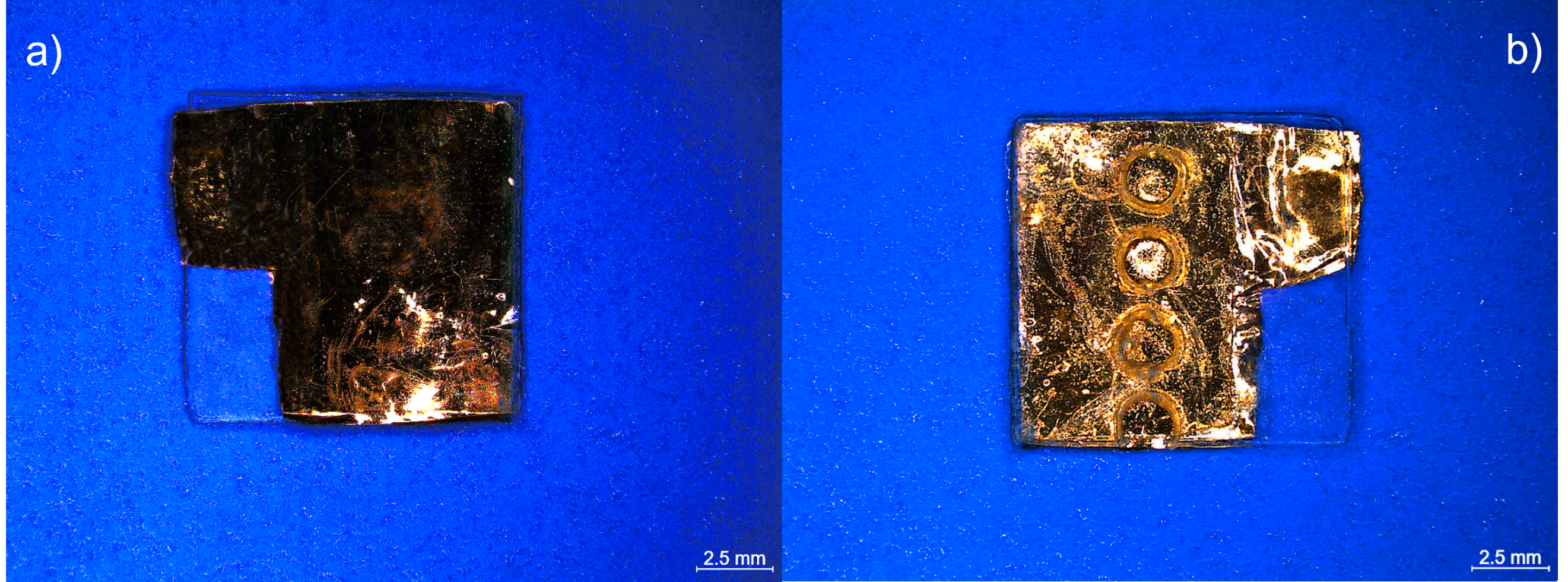}}
\caption{a) Optical image of the second design for the harvester b) Bottom view of the same design.}
\label{fig1}
\end{figure}

Also, based on the stress correlation with strain, which is realized by the Young Modulus (equation 2), the higher magnitude for E results in higher voltage output, which is crucial for energy harvesters.
\begin{equation}
\label{eq1}
V = g_{33}ds.
\end{equation}
\begin{equation}
\label{eq2}
V = g_{33}dE\varepsilon.
\end{equation}
In this way, tunability of $\beta$ phase percentage by chemical methods or electrical poling of the film results in higher crystallinity, which determines the young modulus of PVDF-TrFE films. Thus, we utilized FC20 polymer powders with 20 percent mol content of TrFE ($\beta$ phase) to achieve proper flexibility and stiffness. This way, various PVDF-TrFE films with elastic modulus between 0.7-3 GPa can be selected based on the trade-off between the required flexibility and voltage output in a specific energy harvesting application.
\subsection{Fabrication Process}
To validate the fabrication flow, a 3 mm diameter hole on a 10 × 10 mm PET sheet was created using a mechanical cutter. A thin double-sided tape was put on top of a supportive PET cut into a 7 × 10 mm rectangle. Another PET sheet, which PVDF will later replace, was then cut according to design constraints. This layer also had parts shaped as contact pads with 3 × 3 mm dimensions and was attached to the double-sided tape. To fabricate the device, in the first step, plain PET sheets were cut in 10 × 10 mm size with a mechanical cutter device (step 1). The sheets were 140 µm thick and cut with a laser to form 1.5 mm diameter circular cavities (step 2). A 65 µm thick double-sided tape was then put on top of the PET (step 3). The tape was used to fix the PolyK PVDF film on the sheet. In addition, the tape was used as the structural layer for the harvester, moving the piezoelectric layers away from the neutral axis. Fig. 6 shows the fabrication steps used to fabricate the PEH prototypes. Before the functionality tests, voltage output tests under high impact force were done.
\begin{figure}[H]
\centerline{\includegraphics[width=0.7\columnwidth]{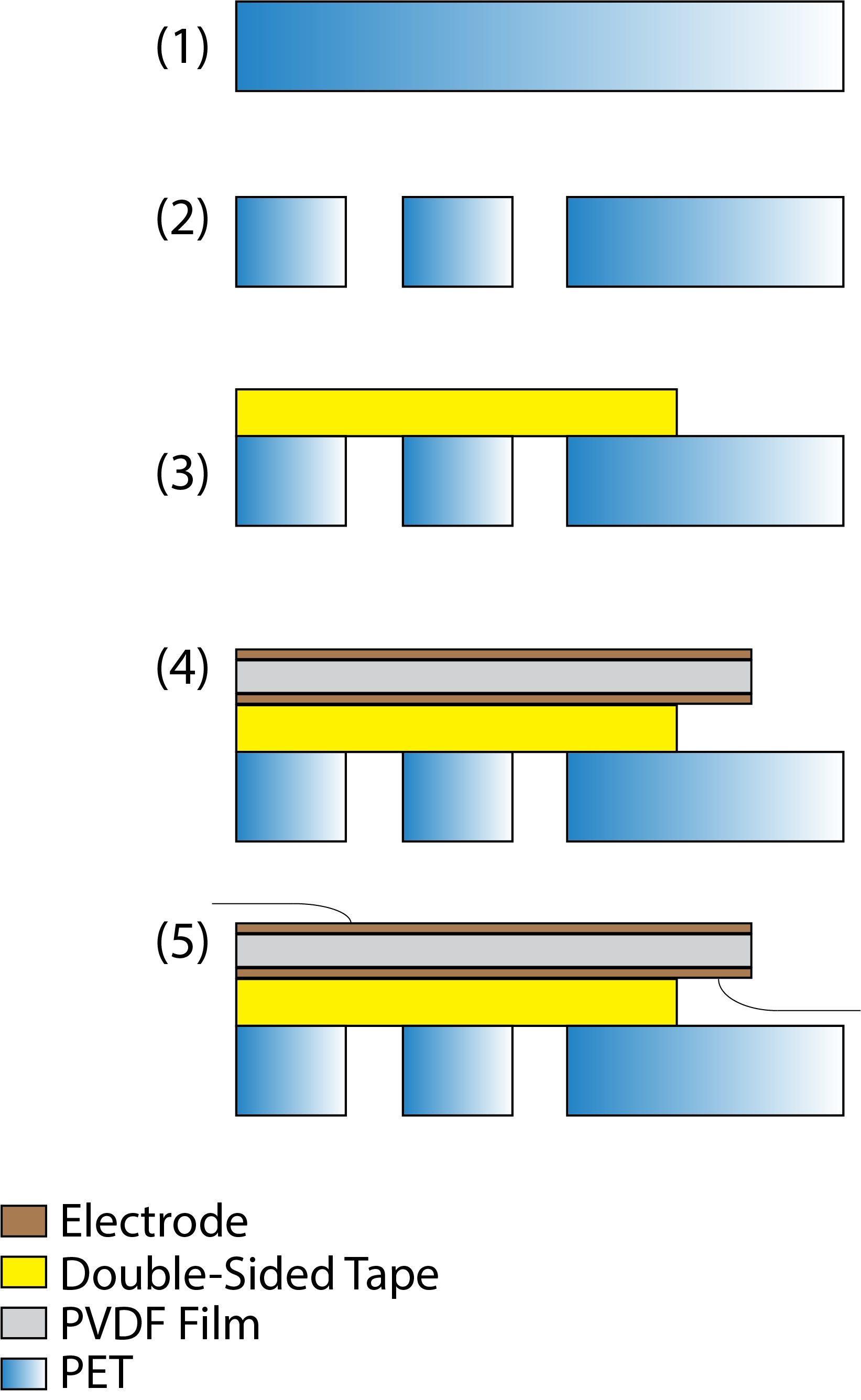}}
\caption{(1) Plain PET sheets were cut in 10 × 10 mm size with
a mechanical cutter device (2) Sheets were cut with a laser to
form 1.5 mm diameter circular cavities (3) A thin double-sided tape was put on top of the PET (4) The tape was used to
fix the piezoelectric film on the sheet (5) Bonding connections
were realized.}
\label{fig1}
\end{figure}

An 18 µm PVDF-TrFE layer was placed on the double-sided tape (step 4). Two 3 × 3 mm electrode pads were cut out of the PVDF film to make bonding easier while getting connections from the device. Connections are made using flat tip connectors (step 5). The sheet was metal coated on both sides. Using the described process, circular plate energy harvesters with 1.5 mm diameter were fabricated. The acoustic resonator tube length was fixed to 140 µm in all prototypes.

\subsection{Tests}
The experiment platform was built as shown in Fig. 7. The setup consisted of a pulse width modulation (PWM) 4wire fan, an Arduino Uno, a dry air supply, and a BMP280 pressure sensor breakout board from Adafruit. The output properties of the resonant cavity piezoelectric energy harvester were studied experimentally. A code was developed to measure the pressure dynamically from serial pins. Serial Peripheral Interface (SPI) protocol was used for communication. A unique test setup was built using 3D printed parts. Dry air was flown towards the fan’s turning propeller, creating pressure fluctuations by creating intervals. The frequency of these intervals was set by changing the PWM frequency of the fan and operating voltage value. PEH was fixed close to the fan to be vibrated by these fluctuations, creating alternating voltage. The maximum wind speed measured was 20 m/s. The rotational speed of the fan was validated using a tachometer. The current output of the harvester was measured using an oscilloscope. The output voltage was converted to the current by dividing the peak-to-peak CH2 output of the oscilloscope with the resistance value (1.5 k$\Omega$).

\begin{figure*}[h]
\centerline{\includegraphics[width=0.9\textwidth]{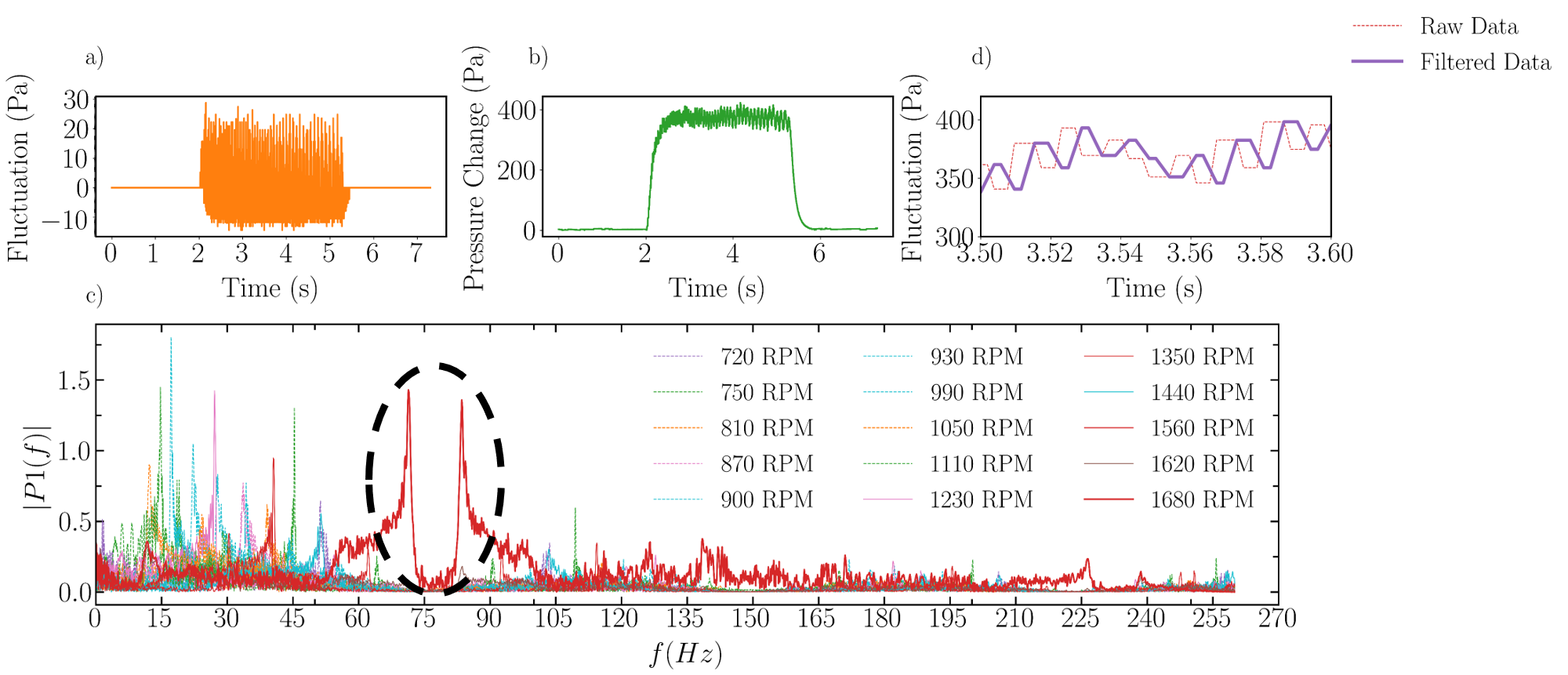}}
\caption{a) Pressure fluctuation spikes were plotted for 1680 RPM measurement showing peak-to-peak 45 Pa pressure changes
per hit b) Using the dry air supply, 400 Pa of pressure could apply to the piezoelectric plates c) Sweeping through 720 RPM to
1680 RPM tests using the PWM fan, FFT was done to each test data extracting the dominant fluctuation frequencies. 1680 RPM
was chosen for the application for having the highest fluctuation frequency having dominant peaks at 72 Hz and 84 Hz d) Testing
data was filtered through moving median filter by taking the average of last 5 values. The 0 Hz noise in the FFT plots were
eliminated using filtering.}
\label{fig1}
\end{figure*}

\begin{figure}[H]
\centerline{\includegraphics[width=.7\columnwidth]{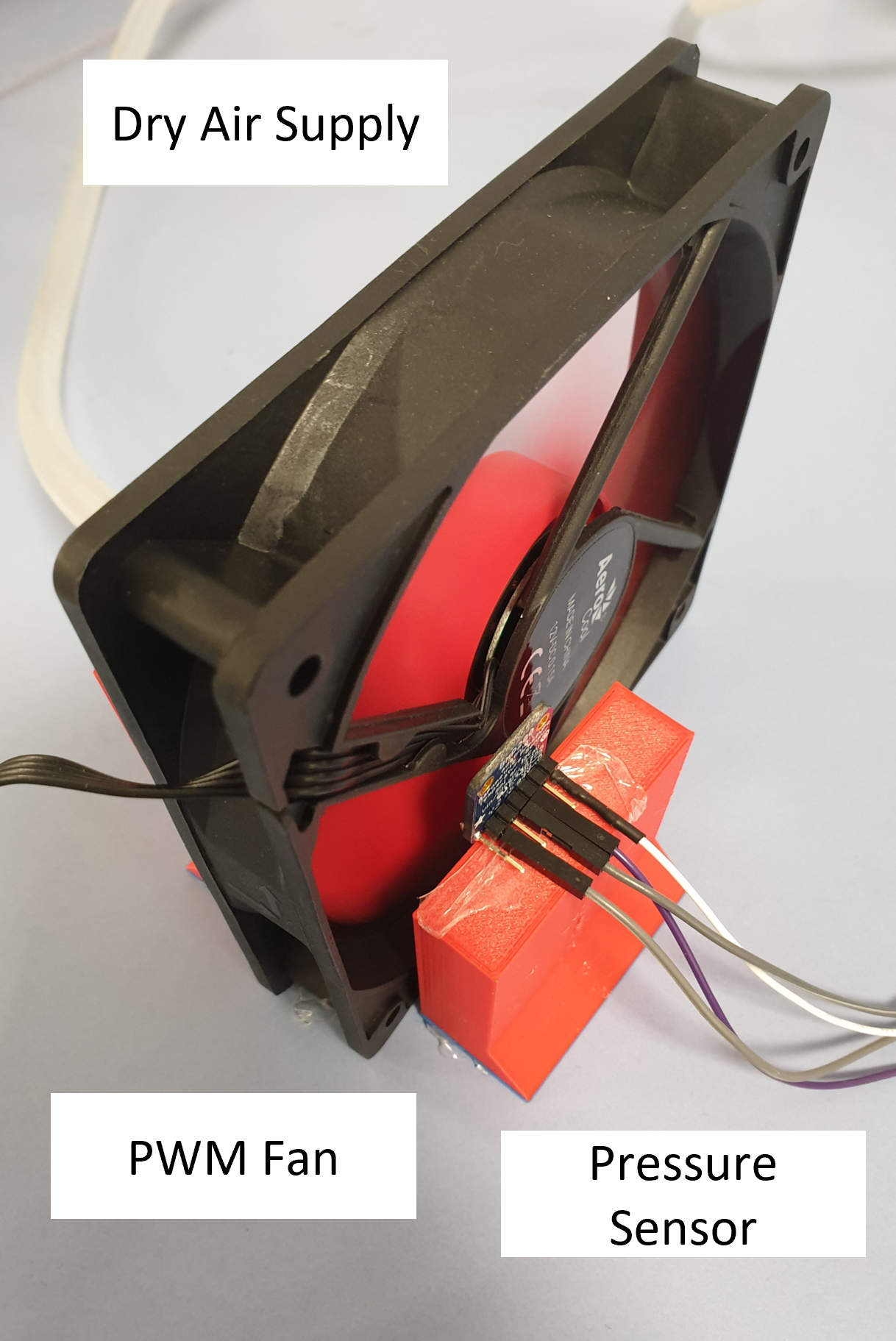}}
\caption{Test setup was built using 3D printed parts which
have a cavity for dry air supply and a surface to fix the pressure
sensor.}
\label{fig1}
\end{figure}

The procedure was applied to both harvester designs.

\section{Results and Discussion}
The research presents two 10 × 10 mm transducers built on the PET layer. The neutral axis and its distance from the midplanes of piezoelectric layers were determined using the measured layer thicknesses; z was calculated as 74 µm. Fig. 8 and Fig. 9 represent the voltage output of the harvester having the first design and characterization plots for the wind setup respectively.
\begin{figure}[H]
\centerline{\includegraphics[width=.9\columnwidth]{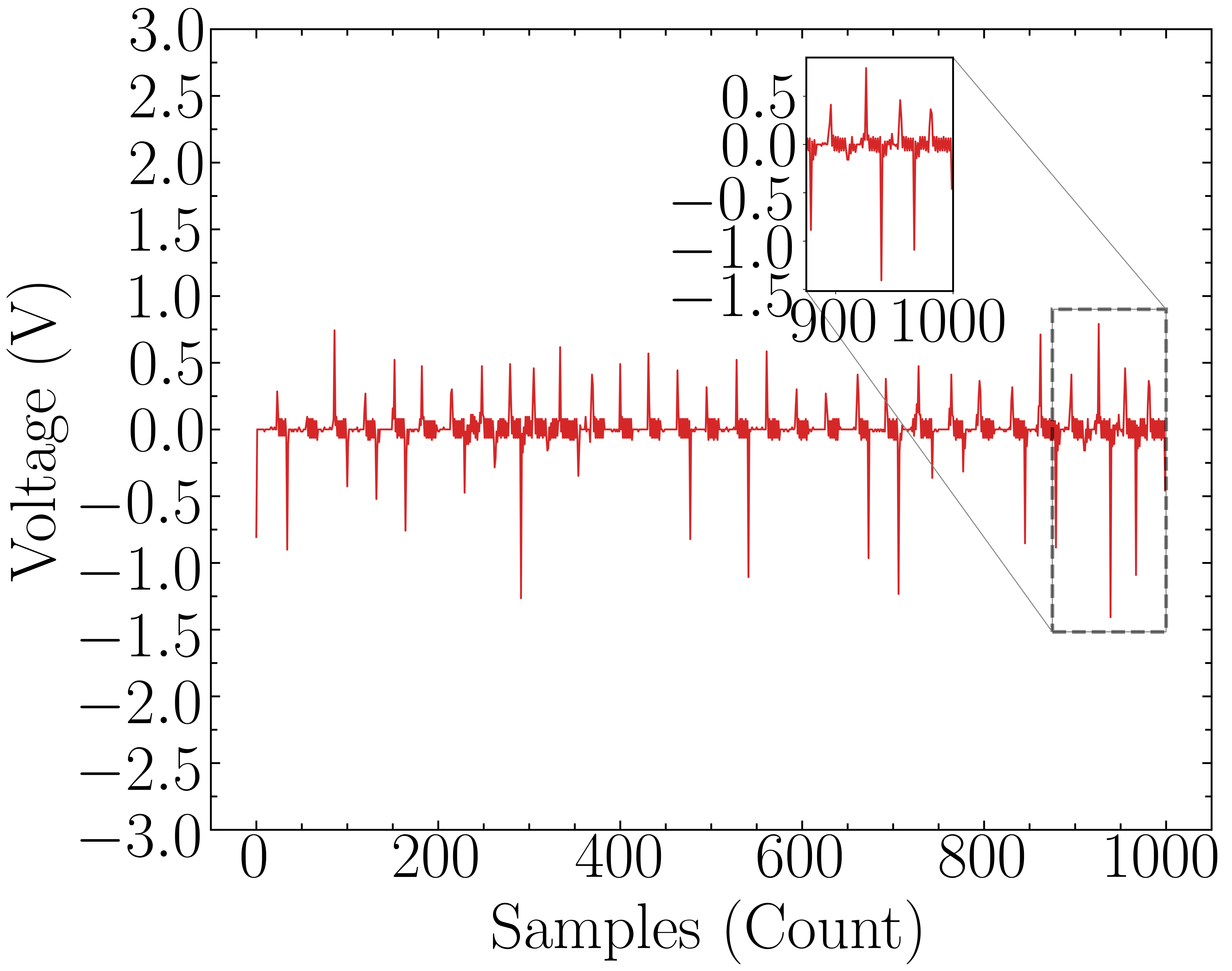}}
\caption{The maximum peak-to-peak value seen was 1.4 V
under impulse (First Design).}
\label{fig1}
\end{figure}

It is also worth noting that the piezoelectric material can withstand significant degrees of strain. Fig. 10 (a-c) demonstrates the harvester’s current and voltage output for two different designs, and the tested harvester is given in Fig. 11.

\begin{figure}[H]
\centerline{\includegraphics[width=\columnwidth]{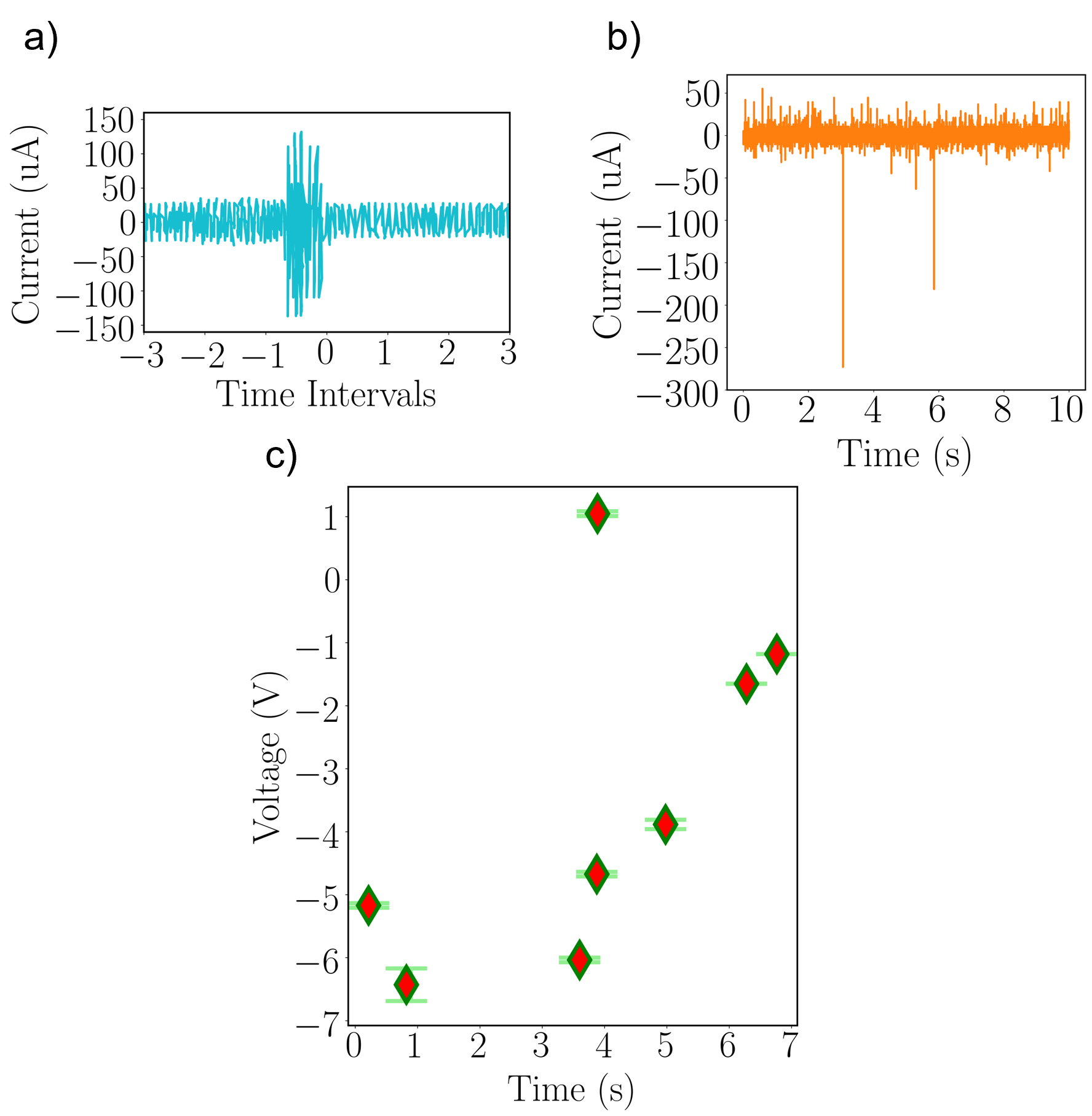}}
\caption{a) The maximum current was 136.9 µA under impulse when taking the average of the peak values (First Design) b) The output current from the sensor under impulse current of  $I_{max}$ = 273 µA (Second Design) c) The output voltage measured from the sensor (Second Design) plotted by taking average values of 5 different impulse tests reporting in error bars. The results are not time dependent. It is worth noting that the output voltage produced is 6.61 $V_{max}$.}
\label{fig1}
\end{figure}

Fig. 23 shows the peak-to-peak voltage graph obtained from
the second design. It is worth noting that the output voltage
produced is 6.61 $V_{max}$.

\begin{figure}[H]
\centerline{\includegraphics[width=0.9\columnwidth]{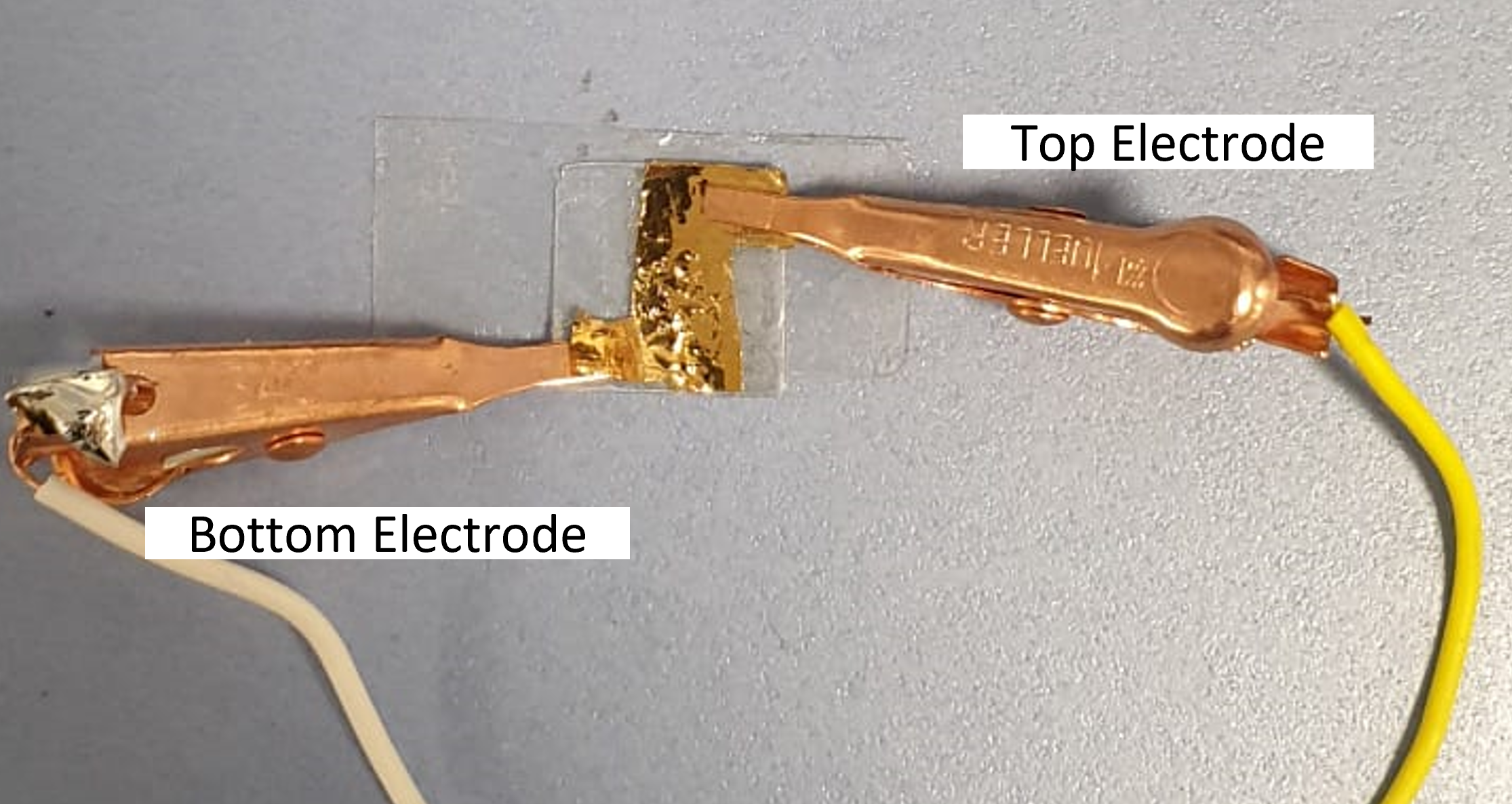}}
\caption{Two flat tip crocodile wires were used to get contacts from the bottom and top electrodes.}
\label{fig1}
\end{figure}

\begin{figure*}[h]
\centerline{\includegraphics[width=.8\textwidth]{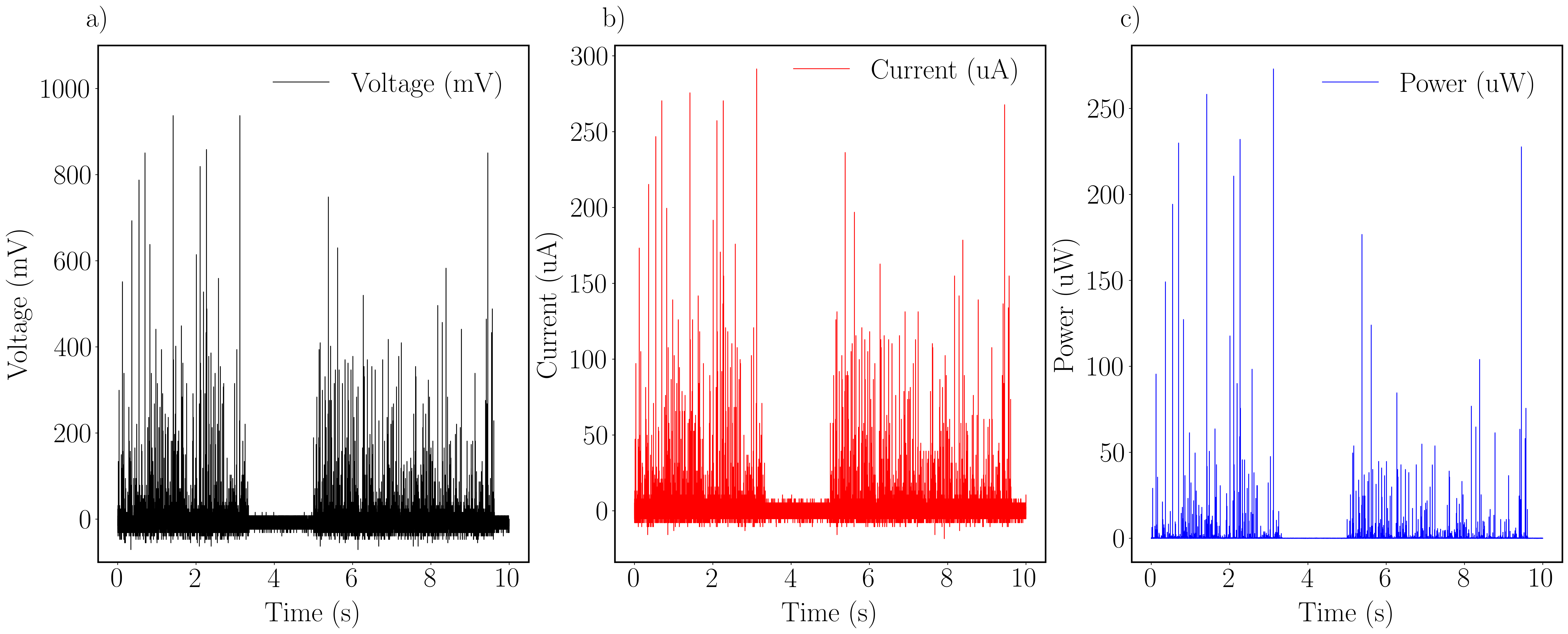}}
\caption{a) The maximum voltage obtained from the wind setup was $V_{max}$ = 937.01 mV while the sensor was connected to the
current measurement circuit b) The maximum current obtained from the wind setup was $I_{max}$ = 291.34 $\mu$A c) The maximum power
obtained from the wind setup was $P_{max}$ = 272.99 $\mu$W.}
\label{fig1}
\end{figure*}

Having higher power output, the second harvester design was evaluated under a wind test setup. Fig. 12 (a-c) demonstrates the harvester's voltage, current, and power output on the same time scale. The power output difference between the two designs is mainly based on the effective piezoelectric layer area, where the second design has a 36.2$\%$ larger piezoelectric layer area than the first design. Impact applied to the sensor for measuring a capacitor charge discharge curve. A full bridge rectifier circuit was used for converting the AC output and initiating capacitor charging. The ceramic capacitor was charged up to 585 mV. The maximum voltage value measured across the capacitor was 760 mV. The voltage increase for a single impulse was 174 mV. The data for capacitor charging is given in Fig. 13. After the force was released from the piezoelectric element, the capacitor charge was discharged due to the 1 M$\Omega$ oscilloscope input impedance. This experiment runs for 10k seconds to show the stability of the capacitor charge.

\begin{figure}[H]
\centerline{\includegraphics[width=\columnwidth]{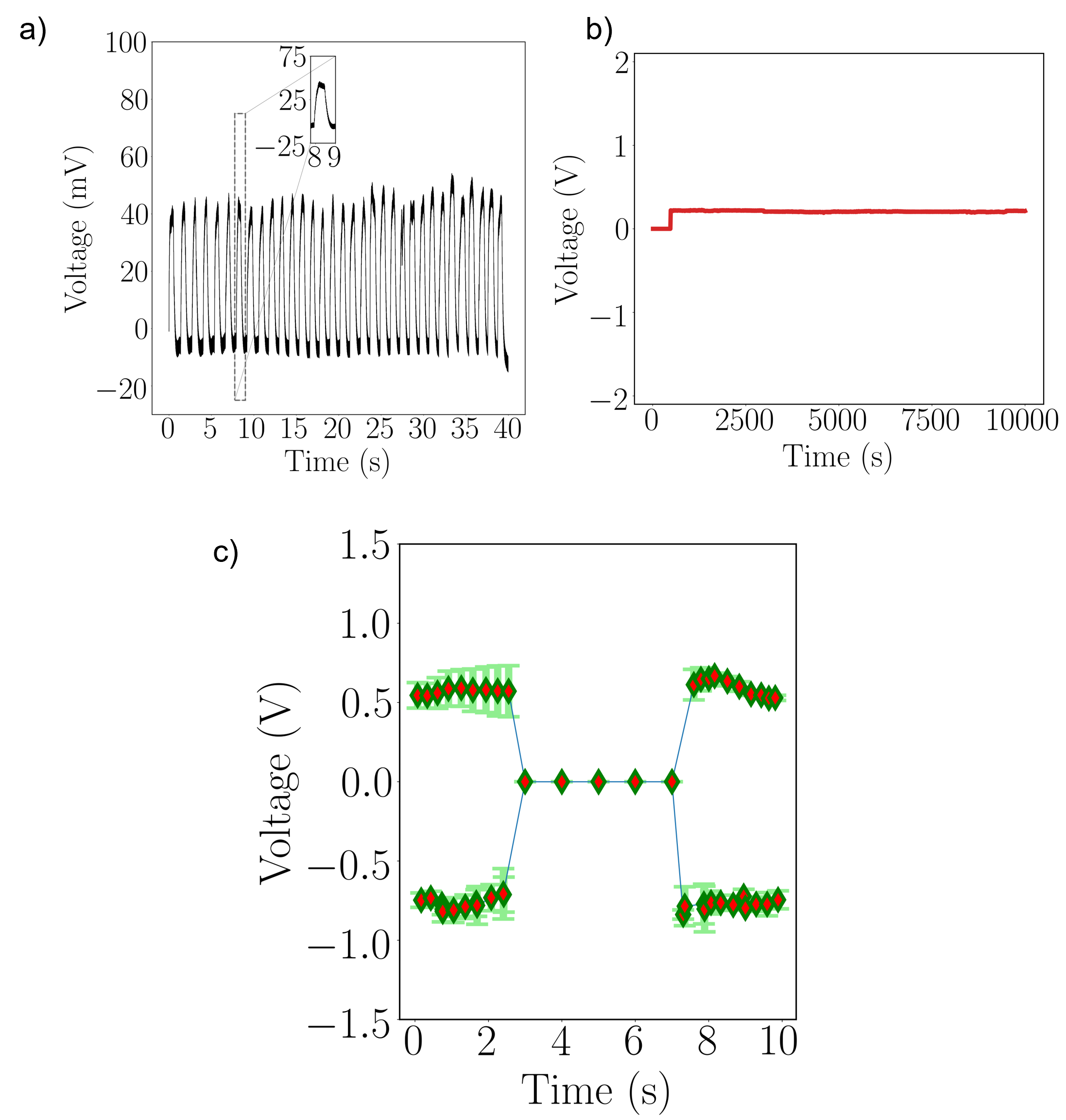}}
\caption{a) Charge of the capacitor was periodically changed due to impulses on the piezoelectric element b) The 104 nF capacitor had a stable charge during the acquisition c) Sensor’s voltage output under wind load without any external circuit connection to validate the values taken from the current measurement circuit.  $V_{pp}$ = 1.74 V was measured from the wind tests.}
\label{fig1}
\end{figure}

The PVDF-TrFe energy harvester can run a MEMS BMP280 (Bosch) pressure sensor widely used in structural health monitoring applications in wind turbines. Needing 2.7 µA current and 1.20 V voltage input, a BMP280 sensor can be run using the PVDF-TrFE energy harvester. Fig. 14 reports the power and current requirements of two other commercial pressure sensors having operating voltages of 1.5 V maximum together with BMP280, showing that the PVDF-TrFe piezoelectric harvester can power these devices.

\begin{figure}[H]
\centerline{\includegraphics[width=\columnwidth]{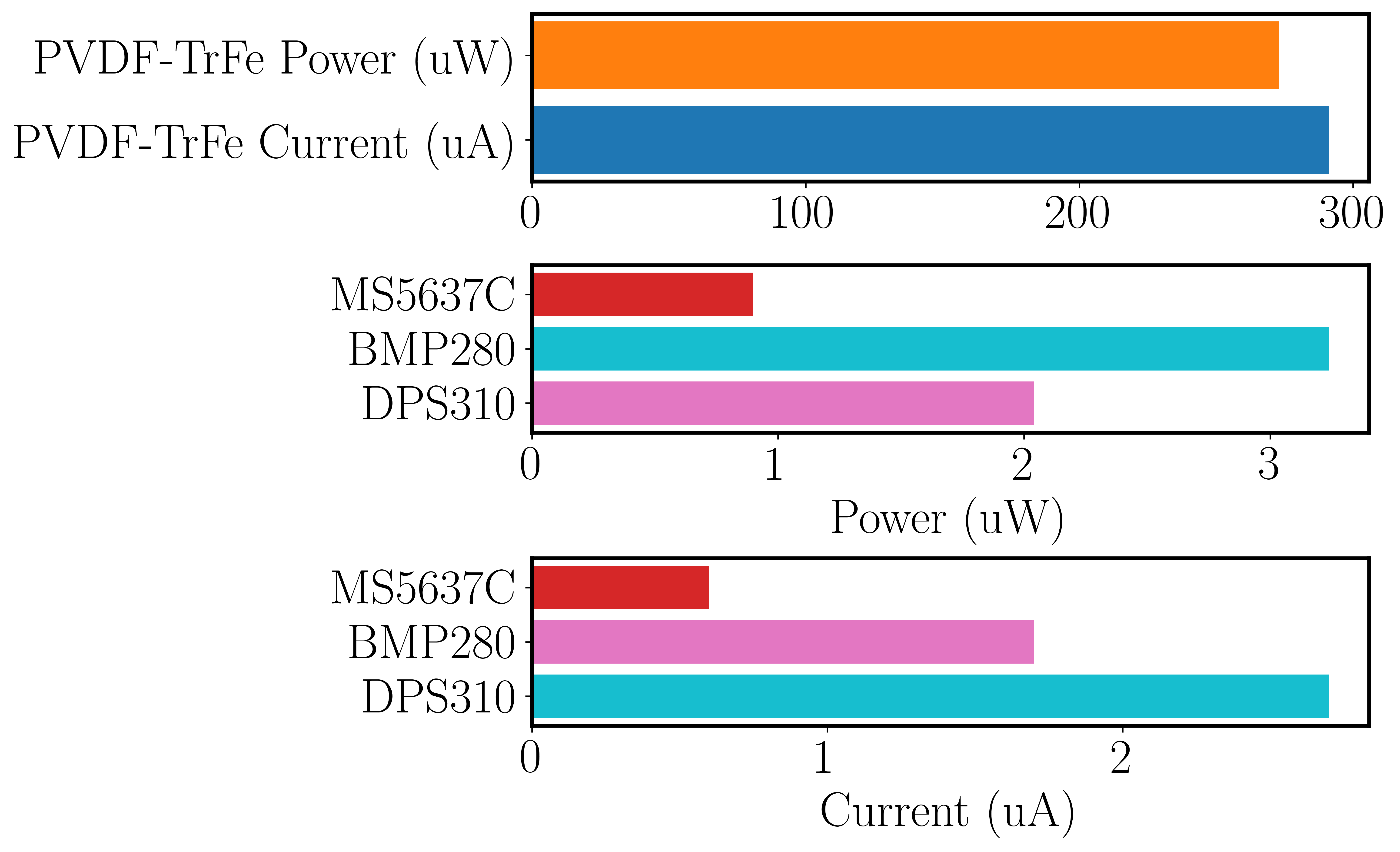}}
\caption{MS5637C (MEAS Switzerland), DPS310 (Infineon),
and BMP280 (Bosch) are widely used commercial pressure
sensors that can be powered with the harvester output [26-28].}
\label{fig1}
\end{figure}

\section{Conclusion}

PVDF-TrFE has significant advantages over other ferroelectric
materials due to its soft and flexible character. This study uses
conventional fabrication techniques to describe the design,
construction, and characterization of PEHs with PVDF-TrFE
layers. The proposed PEH structures are thin films with circular
plates of 1.5 mm. The device's dimensions were chosen to keep
the neutral axis outside the active piezoelectric layers. Two
sensor designs were used in this experiment. The thickness of
the PEH designs was 223 µm after fabrication. On an impulse
load, the maximum harvested voltage was measured at 6.61 V
and the current measured at 273 µA; on a wind load, the
maximum harvested voltage was measured at 937.01 mV and
the current measured at 291.34 µA giving a maximum power
value of 272.99 µW for the second energy harvester design.
\section*{Acknowledgment}
The authors thank BMDL lab members at Koç University for
their support in device fabrication and characterization,
Muhammad Junaid Akhtar for current measurement trials, and
Berke Ataseven for FFT code development.

\begin{IEEEbiography}[{\includegraphics[width=1in,height=1.25in,clip,keepaspectratio]{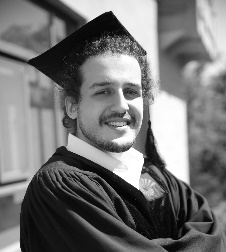}}]{Berkay Kullukçu} is a master's student in the Mechanical Engineering Department at Koç University. His research topic focuses on piezoelectric PVDF-TRFE/ PET energy harvesters for structural health monitoring (SHM) applications. The aim of his research is to develop a flexible energy harvester for powering low power electronics. He obtained his BSc in Mechanical Engineering from Koç University in 2020.
\end{IEEEbiography}
\begin{IEEEbiography}[{\includegraphics[width=1in,height=1.25in,clip,keepaspectratio]{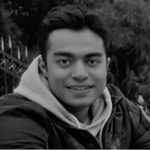}}]{Muhammad J. Bathaei} is a graduate student in Biomedical Engineering and started his PhD in February 2020. He received his BSc degree in Materials Science and Engineering from Tehran University (UT). His research currently focuses on biodegradable sensors for medical applications under supervision of Prof. Levent Beker at Bio-integrated Microdevices laboratory. 
\end{IEEEbiography}
\begin{IEEEbiography}[{\includegraphics[width=1in,height=1.25in,clip,keepaspectratio]{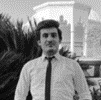}}]{Muhammad Awais} is a Ph.D. student in the Biomedical engineering department of Koç University. Currently, he is working on wearable ultrasonic sensors and implantable electronics for biomedical purposes. He obtained his bachelor's degree in Electronics engineering from Ghulam Ishaq Khan Institute of Engineering Sciences and Technology Topi, Pakistan.
\end{IEEEbiography}
\begin{IEEEbiography}[{\includegraphics[width=1in,height=1.25in,clip,keepaspectratio]{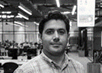}}]{Hadi Mirzajani} is pursuing his research as a postdoc at the Mechanical Engineering Department of Koç University. His focus is on wireless sensing patches for noninvasive or minimally invasive diagnosis of disease. His topics of interest are biosensors and bioelectronics, microfluidics, wireless power/data transmission, and RF MEMS. He has published more than 25 scientific papers and served as reviewer for many journals and conferences.
\end{IEEEbiography}
\begin{IEEEbiography}[{\includegraphics[width=1in,height=1.25in,clip,keepaspectratio]{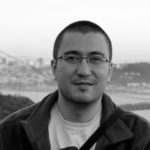}}]{Levent Beker} got undergrad and master's degrees in mechanical engineering and micro-nano technology from the Middle East Technical University (METU), Ankara, Turkey.
Then, he got his PhD in mechanical engineering from the University of California, Berkeley, U.S. In 2017, he joined Prof. Zhenan Bao's group at Stanford University and in September 2019, he joined Koç University as an assistant professor and established the Bio-integrated microdevices laboratory.

\end{IEEEbiography}

\end{document}